\documentclass{article}
\usepackage{waspaa19,amsmath,graphicx,url,times}
\usepackage{booktabs}
\usepackage{graphicx}
\usepackage{color}
\usepackage{lipsum}
\usepackage{subfig}
\usepackage{comment}
\usepackage{multirow}
\usepackage[normalem]{ulem}
\usepackage{textcomp}
 
\title{Sound Search by Text Description or Vocal Imitation?}

\name{Yichi Zhang, Yiting Zhang, and Zhiyao Duan\thanks{This work is funded by the National Science Foundation grant No. 1617107. We acknowledge NVIDIA's GPU donation for this research.}}
\address{Dept. of Electrical and Computer Engineering, University of Rochester, Rochester, NY 14627, USA\\
       \emph{\{yichi.zhang, yiting.zhang, zhiyao.duan\}@rochester.edu}}

\begin{document}

\ninept
\maketitle

\begin{sloppy}

\begin{abstract}
  Searching sounds by text labels is often difficult, as text descriptions cannot describe the audio content in detail. Query by vocal imitation bridges such gap and provides a novel way to sound search. Several algorithms for sound search by vocal imitation have been proposed and evaluated in a simulation environment, however, they have not been deployed into a real search engine nor evaluated by real users. This pilot work conducts a subjective study to compare these two approaches to sound search, and tries to answer the question of which approach works better for what kinds of sounds. 
  To do so, we developed two web-based search engines for sound, one by vocal imitation (\emph{Vroom!}) and the other by text description (\emph{TextSearch}). We also developed an experimental framework to host these engines to collect statistics of user behaviors and ratings. Results showed that \emph{Vroom!} received significantly higher search satisfaction ratings than \emph{TextSearch} did for sound categories that were difficult for subjects to describe by text. Results also showed a better overall ease-of-use rating for \emph{Vroom!} than \emph{TextSearch} on the limited sound library in our experiments. These findings suggest advantages of vocal-imitation-based search for sound in practice.
\end{abstract}


\section{Introduction}
\label{sec:intro}

Designing methods to access and manage multimedia documents such as audio recordings is an important information retrieval task. Traditional search engines for audio files use text labels as queries. However this is not always effective. First, it requires users to be familiar with the audio library taxonomy and text labels, which is unrealistic for many users with no or little audio engineering background. Second, text descriptions or metadata are abstract and do not describe the audio content in detail. Third, many sounds, such as those generated by computer synthesizers, lack commonly accepted semantic meanings and text descriptions. 

Vocal imitation is commonly known as using voice to mimic sounds. It is widely used in our daily conversations, as it is an effective way to convey sound concepts that are difficult to describe by language. For example, when referring to the "Christmas tree" dog barking sound, vocal imitation is more intuitive compared to text descriptions. Hence, designing computational systems that allow users to search sounds through vocal imitation \cite{dBlancas2014,yZhang2015} goes beyond the current text-based search and enables novel human-computer interactions. It has natural advantages over text-based search as it does not require users to be familiar with audio taxonomy and it indexes the detailed audio content instead of abstract and often non-agreeable text descriptions. Regarding applications, sound search by vocal imitation can be found useful in many fields including movie and music production, multimedia retrieval, and security and surveillance. 

In our previous work, we proposed a deep neural network model called TL-IMINET \cite{yZhang2018} for sound search by vocal imitation. It addresses two main technical challenges: 1) feature learning: what feature representations are appropriate for the vocal imitation and reference sound, and 2) metric learning: how to design the similarity between a vocal imitation and each sound candidate. Experiments on the VocalSketch Data Set \cite{mCartwright2015} have shown promising retrieval performance of this model, however, no user studies have been conducted to validate the model and the sound-search-by-vocal-imitation approach in general at the system level. As a follow-up study, several questions naturally arise: 1) Is vocal-imitation-based search an acceptable approach to sound search for ordinary users without an extensive audio engineering background? 2) How does vocal-imitation-based search compare with the traditional text-based search for different kinds of sounds in terms of search effectiveness and efficiency? 



To answer the above questions, in this work, we conduct a subjective study to compare sound search by vocal imitation and by text description. Specifically, we designed a web-based search engine called \emph{Vroom!}. 
It allows a user to record a vocal imitation as a query to search sounds in a small sound library using a pre-trained TL-IMINET model as the search algorithm. We also designed another web-based search engine called \emph{TextSearch}. It allows a user to search sounds using keywords. To compare the two systems, we recruited 23 subjects to search for a randomized list of 20 sounds in a small sound library containing three different sound categories using both systems. We built a web-based experimental framework to collect statistics of search behaviors and ratings. Analyses of the results show that subjects gave significantly higher overall ease-of-use scores to \emph{Vroom!} than \emph{TextSearch} in this sound library. Results also show significant advantages of \emph{Vroom!} over \emph{TextSearch} on categories that were difficult to describe by text.

\section{Related Work}
\label{sec:format}

Sound search by text description has been widely accepted in our daily life. For example, Freesound \cite{freesoundweb} is an online collaborative sound database with more than 400,000 sounds. Those sounds are tagged with text descriptions for text-based search. SoundCloud \cite{soundcloudweb} is another online audio distribution platform that enables users to search sounds by text description. 

On the other hand, sound search by vocal imitation is drawing increasing attention from the research community to address limitations of text-based search. It belongs to the task of Query by Example (QBE) \cite{mZloof1977}. There are numerous QBE applications in the audio domain, such as audio fingerprinting of the exact match \cite{aWang2003} or live versions \cite{zRafii2014, tTsai2016}, cover song detection \cite{tBertin2011} and spoken document retrieval \cite{tChia2008}. 
Vocal imitation of a sound has been first proposed for music retrieval, such as finding songs by humming the melody as a query \cite{aGhias1995, dannenberg2007comparative} or beat boxing the rhythm \cite{aKapur2004, oGillet2005}. Recently, it has been extended for general sound retrieval, as summarized below. 

Roma and Serra \cite{gRoma2015} designed a system that allows users to search sounds on Freesound by recording audio with a microphone, but no formal evaluation was reported. Blancas et al. \cite{dBlancas2014} built a supervised system using hand-crafted features by the Timbre Toolbox \cite{gPeeters2011} and an SVM classifier. 
The major limitation of supervised systems, however, is that they cannot retrieve sounds that do not have training imitations. Hel\'en and Virtanen \cite{mHelen2010} designed a query by example system for generic audio. Hand-crafted frame-level features were extracted from both query and sound samples and the query-sample pairwise similarity was measured by probability distribution of the features. 

In our previous work \cite{yZhang2015}, we first proposed a supervised system using a Stacked Auto-Encoder (SAE) for automatic feature learning followed by an SVM for imitation classification. We then proposed an unsupervised system called IMISOUND \cite{yZhang2016} to overcome the close-set limitation in \cite{yZhang2015}. The SAE was adopted for feature extraction for both imitation queries and sound candidates and various similarity measures were calculated \cite{sKullback1951, hSakoe1978, Dehak10}. 
Due to the separation of feature representation and metric learning, we further proposed the end-to-end Siamese style convolutional neural networks \cite{yZhang2017} to integrate these two modules together, in which the transfer learning based TL-IMINET is our most updated model \cite{yZhang2019J}. Meanwhile, the benefits of applying positive and negative imitations to update the cosine similarity between the query and sound candidate embedding was investigated in \cite{bKim2019}. To understand what such neural networks actually learns, we also visualized and sonified the input patterns in TL-IMINET \cite{yZhang2018} using activation maximization \cite{dErhan2010}.


Up to date, research on sound search by vocal imitation has been only conducted at the algorithm development level. No usable search engines have been deployed based on these algorithms, nor user studies have been conducted to assess the effectiveness of the new search approach in practice. This paper conducts a pilot study along this line: comparing a text-based search engine with a vocal-imitation-based search engine built on the TL-IMINET algorithm.

\section{Search Engines for Comparison}
\subsection{Search by Vocal Imitation: \emph{Vroom!}}
\label{sec:vroom}

We designed a web-based sound search engine by vocal imitation, called \emph{Vroom!}. 
The frontend GUI is designed using Javascript, HTML, and CSS languages. It allows a user to record a vocal imitation of sound that he/she is looking for using the recorder.js Javascript library \cite{recorderjsweb}. It also allows the user to listen to the recording, inspect the waveform, and re-record imitations. To search, the user specifies a sound category and clicks on the ``Go Search!'' button.
The recording is then uploaded to the backend server to compare with each sound within the specified category using the TL-IMINET algorithm. Top five sound candidates with the highest similarity scores are returned to the user. The user can listen to the returned sounds and make a selection to complete the search. If not satisfied with any of the returned sounds, the user can re-record an imitation and re-do the search. The GUI is shown in Figure~\ref{Fig_VocalTextComparison}(a).


\begin{figure}[!t]
\centering
\includegraphics[width=\columnwidth]{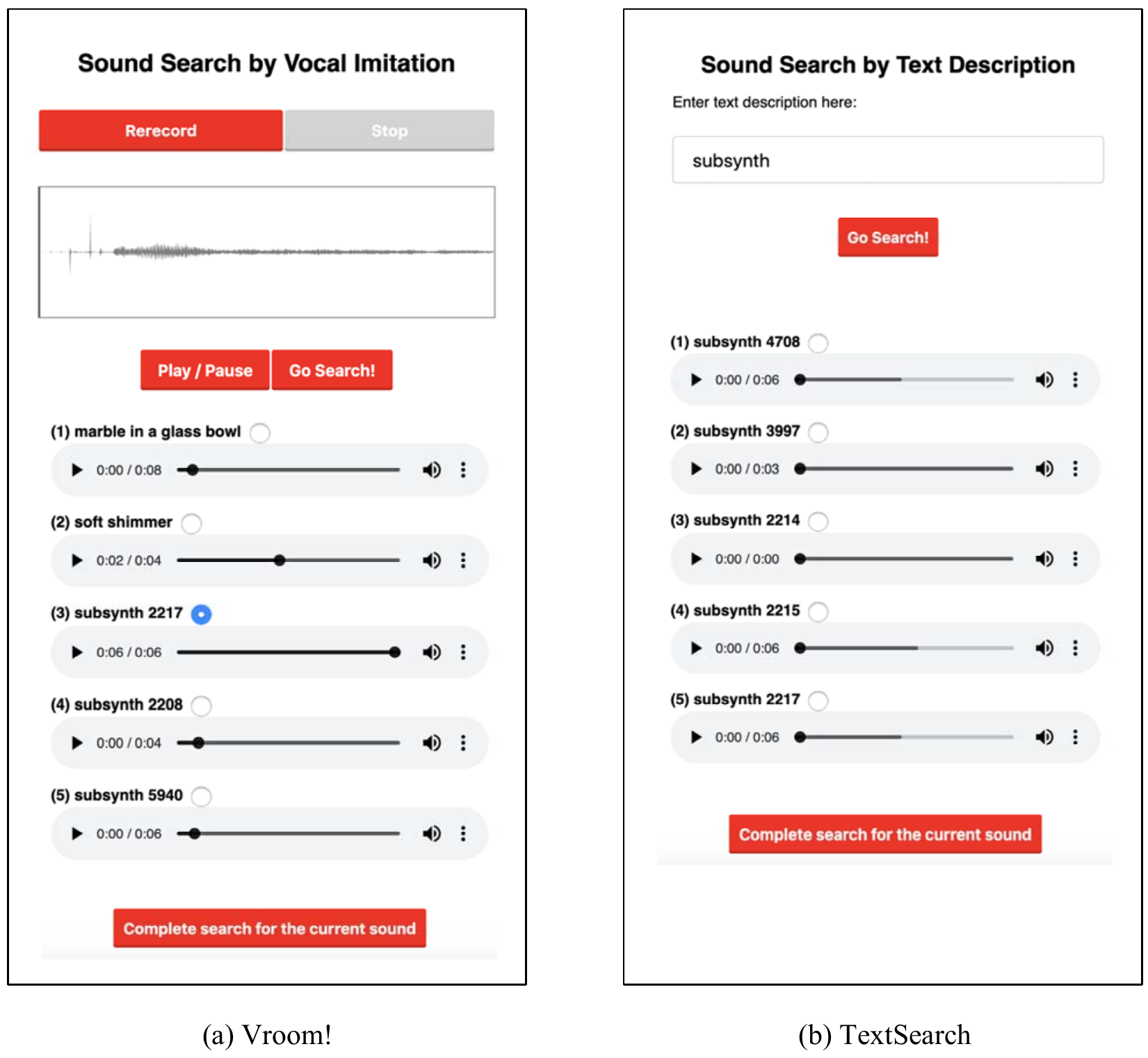}
\caption{Frontend GUIs of (a) the vocal-imitation-based search engine \emph{Vroom!} and (b) the text-based-search engine \emph{TextSearch}.}
\label{Fig_VocalTextComparison}
\end{figure}

The backend is hosted on a Ubuntu system installing Keras v2.2.4 with GPU acceleration. It implements the TL-IMINET model in the Node.js express framework. It responds to each frontend search request from users and retrieves the most similar sounds to each imitation, within the specified sound category of a pre-defined sound library.

The overall structure of TL-IMINET \cite{yZhang2019J} contains two Convolutional Neural Network (CNN) towers for feature extraction: One tower receives a vocal imitation (the query) as input. The other receives a sound from the library (the candidate) as input. Each tower outputs a feature embedding. These embeddings are then concatenated and fed into a Fully Connected Network (FCN) for similarity calculation. The final similarity output of the Siamese-style neural network is the probability of being a positive pair between the query and the candidate. The CNNs and FCN are trained jointly on positive (i.e., related) and negative (i.e., non-related) query-candidate pairs. Through joint optimization, feature embeddings learned by the CNNs are better tuned for the FCN's metric learning.

The recording and imitation towers for feature extraction are adapted from environmental sound classification \cite{jSalamon2017} and spoken language recognition \cite{gMontavon2009} tasks, respectively. The two tower weights and biases are initialized by pre-training them on external datasets for these tasks. They are then fine-tuned together with the FC layers on the sound retrieval task. 


\subsection{Search by Text Description: \emph{TextSearch}}
\label{sec:textsearch}

We also designed a comparative web-based sound search engine by text description. As shown in Figure~\ref{Fig_VocalTextComparison}(b), the GUI provides the user with a text box to enter keywords that are not case sensitive. The backend server receives the keywords and looks up the entire WordNet dictionary \cite{gMiller1995} online via a Python interface to find synonyms of that query. Then the synonyms together with the query itself are used to search for any matched record, which are then all returned to the user and displayed in the order as stored in the library. Note that if the input keyword is a phrase instead of a single word, synonyms of the entire phrase is searched for. Subjects were notified that single word per search is preferable to return more sound candidates although phrases were also allowed.


\begin{figure*}[!t]
\centering
\includegraphics[width=5.4in]{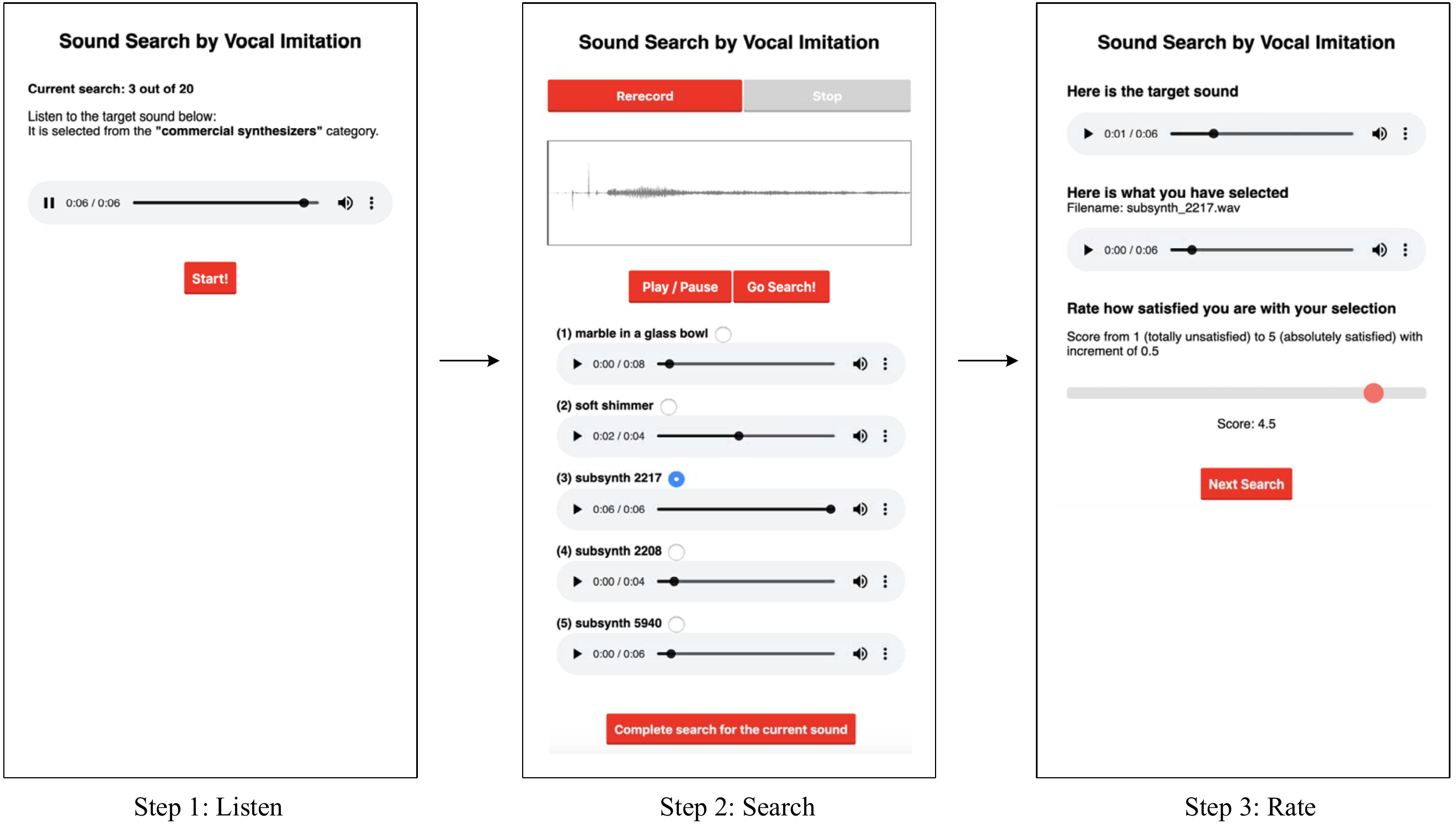}
\caption{Experimental framework hosting the proposed vocal imitation based search engine \emph{Vroom!}. The framework hosting the text description based search engine \emph{TextSearch} is exactly the same but replacing \emph{Vroom!} in Step 2 with \emph{TextSearch}.}
\label{Fig_ExperimentalFramework}
\end{figure*}

\section{Subjective Evaluation}
\label{sec:subjective_evaluation}

\subsection{Experimental Framework}
\label{ssec:framework}
To quantify search behaviors and user experiences and to make quantitative comparisons between \emph{Vroom!} and \emph{TextSearch}, we designed an experimental framework that wraps around each search engine. The experimental framework is another web application. It guides each subject to make 20 searches and rate the ease of use for the search engine after completing all searches. For each search, it guides the subject through three steps. 
In Step 1, the subject listens to a reference sound randomly chosen from a category of the sound library. The category name is visible while the keywords of the sound is not provided. This sound will be the target sound to search in following steps. In Step 2, the reference sound is hidden from the subject, and the subject uses the search engine (\emph{Vroom!} or \emph{TextSearch}) to search for the reference sound in the specified category of the sound library. 
In Step 3, the reference sound appears again. The user compares it with their retrieved sound to rate their satisfaction about the search. These three steps, for the \emph{Vroom!} search engine, are shown in Figure~\ref{Fig_ExperimentalFramework} for illustration.

The experimental framework tries to mimic the search processes in practice as much as possible, however, certain modifications have to be made to allow quantitative analysis. In practice, a user rarely listens to the exact target sound before a search; they usually only have a rough idea about the target sound in their mind to cast their query (imitation or text). In our experimental framework, however, before each search, the subject listens to the target sound to cast their query. While this may positively bias the quality of the query (especially for the imitation query), this is necessary to control what sound to search by the subjects. This is especially important for the small sound library used in this pilot study; the library may simply not contain the target sound if we allowed subjects to search freely. To reduce this positive bias, we hid the target sound during the search (Step 2).


The backend of this experimental framework records statistics of important search behaviors. They include the number of returned sounds played, the number of ``Go Search!'' button clicked, the total time spent, and the rank of the target sound in the returned list for each search. The user satisfaction rating for each search and the ease-of-use rating for the search engine are also collected. 


\subsection{Dataset}
\label{ssec:dataset}

We adopt VocalSketch Data Set v1.0.4 \cite{mCartwright2015} in our experiments. This dataset contains 120 sounds with distinct concepts and 10 vocal imitations for each sound from different Amazon Mechanical Turkers. The sounds and imitations are 3-second long each on average. These sounds are from four categories: Acoustic Instruments (AI), Everyday (ED), Single Synthesizer (SS) and Commercial Synthesizers (CS). The number of sounds in these categories is 40, 120, 40 and 40, respectively. We choose half of the sounds of each category and all of their imitations to compose a dataset to train and validate the backend TL-IMINET model. We use the other half sounds to conduct subjective evaluation. Therefore, training and subjective evaluation materials do not share any sound concepts. During subjective experiments, subjects are asked to search each sound within a specified category. As the differences between SS and CS are difficult to understand by ordinary users, we merge these two categories into one and name it Synthesizers (SN) in our experiments. Therefore the new categories and their sizes in the searchable database are AI (20), ED (60), and SN (40). 

Keywords of these sounds are manually created using their filenames in the dataset. For example, the sound file ``$marimba\_hit\_with\_a\_rubber\_mallet.wav$'' has the following keywords:   ``marimba'', ``hit'', ``rubber'', and ``mallet''. On average each sound associates with around 2 keywords. It is noted that filenames of some sounds in the SN category, i.e., those belonging to the Single Synthesizer category in the VocalSketch dataset, only contain synthesizer parameters instead of semantic words; this makes them essentially not searchable by text. We could have annotated these sounds with some semantic words, however, we did not do so for two reasons: 1) such sounds without semantic keywords do exist in practice, and 2) most of these sounds simply do not have commonly agreeable text descriptions to annotate. 


\subsection{Subjects}
\label{ssec:experimental_setup}

We recruited 23 students with different academic backgrounds from the University of Rochester as our subjects. Each of them was asked to perform 20 sound searches using both \emph{Vroom!} and \emph{TextSearch}, rate their satisfaction score about each search, and rate the ease-of-use score of both systems after completing all 20 searches. Subjects were informed about the collection of their search behaviors and ratings before the experiments. Experiments were conducted in a quiet sound booth and subjects recorded their voice using the built-in microphone of a 2015 Apple MacBook Pro laptop computer. It took about 1 hour for each subject to complete the experiment on average, and each was compensated with a $15$ US dollar gift card. During the experiments, 3 subjects encountered audio play issues in \emph{TextSearch} and could not complete the experiment. We thus excluded their results, resulting in a total of 20 valid subjective experiments.

\subsection{Results}
\label{ssec:results}


Figure \ref{Fig_UserRating} compares two types of user ratings between \emph{Vroom!} and \emph{TextSearch}: 1) User's satisfaction rating (SAT) indicates how satisfied a user is with each search by comparing the finally retrieved sound to the reference sound (collected in Step 3 in Figure \ref{Fig_ExperimentalFramework}); 2) ease-of-use rating evaluates a user's overall experience of each search engine upon the completion of all 20 searches. 

\begin{figure}[!t]
\centering
\includegraphics[width=6.5 cm]{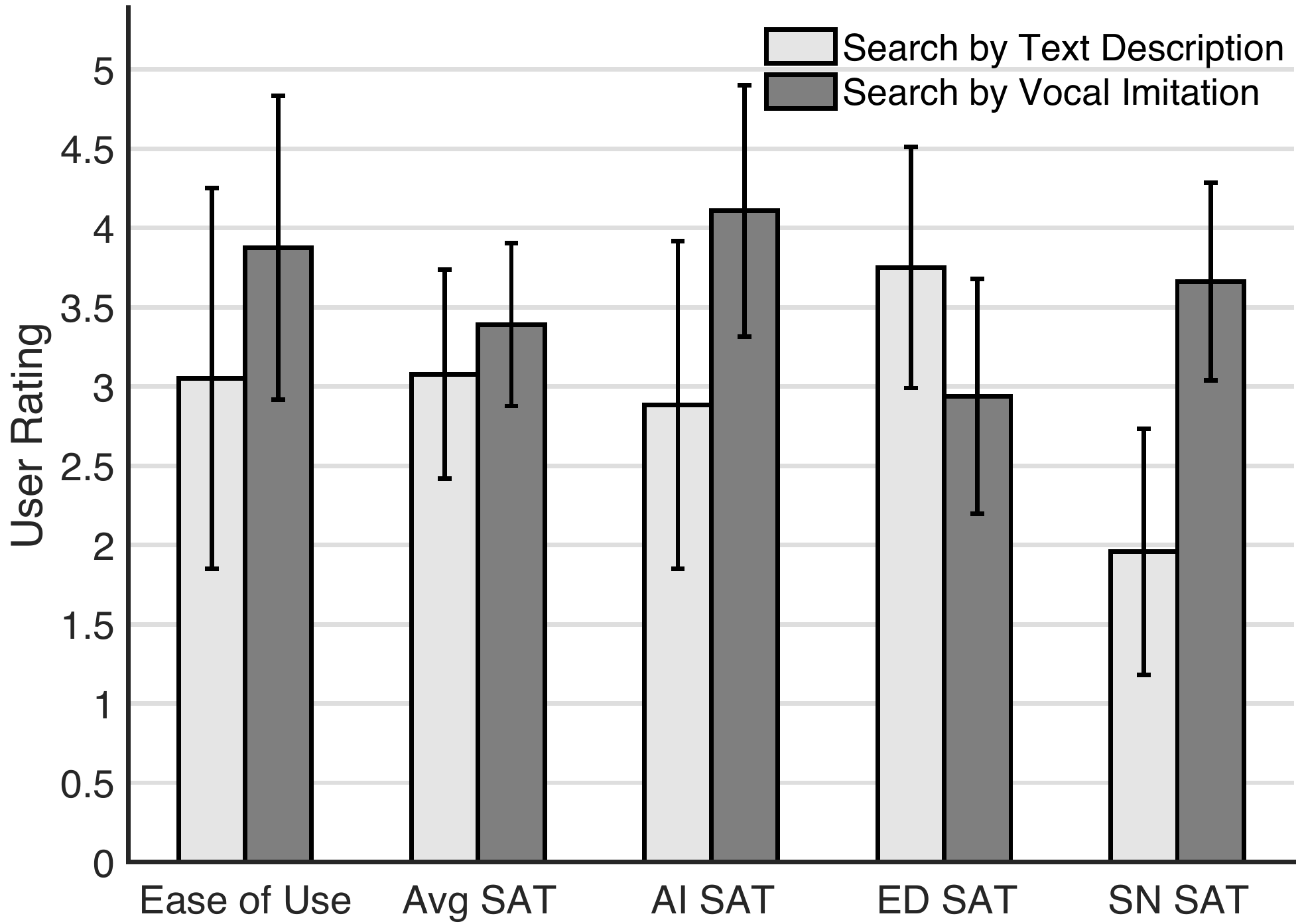}
\caption{Average user ratings of sound search by text description (\emph{TextSearch}) and vocal imitation (\emph{Vroom!}). Ratings include the overall ease-of-use rating of the two search engines, and search satisfaction (SAT) within each sound category and across all categories. Error bars show standard deviations.}
\label{Fig_UserRating}
\end{figure}

It can be seen that \emph{Vroom!} shows a statistically significantly higher ease-of-use rating than \emph{TextSearch} at the significance level of 0.05 (p=0.0108, unpairted t-test). This aligns with the average satisfaction rating of all categories. However, a further inspection reveals that the average satisfaction rating varies much from one category to another. For AI and SN categories, \emph{Vroom!} receives a statistically significantly higher satisfaction rating than \emph{TextSearch} does, at the significance level of 0.01 (AI p = 8.42e-5 and SN p = 2.21e-9, unpaired t-tests). This is because many subjects could not recognize sounds from these categories nor find appropriate keywords to search in \emph{TextSearch}. This was especially the case for the SN category, as many sounds simply do not have semantically meaningful and commonly agreeable keywords. On the other hand, imitating such sounds was not too difficult for many subjects. For the ED category, however, \emph{TextSearch} outperforms \emph{Vroom!}. Subjects were more familiar with these everyday sounds and knew how to describe them with keywords, while some sounds could be difficult to imitate, e.g., shuffling cards and toilet flushing. 




\begin{table}[t]
\centering
\caption{User behavior statistics of \emph{TextSearch} and \emph{Vroom!}.}
\label{mrr_comparisons_symm}
\begin{tabular}{@{}cccc@{}}
\toprule


User Behavior & \emph{TextSearch} & \emph{Vroom!} \\ \midrule
No. search trials & 2.39 $\pm$ 0.84 & 1.65 $\pm$ 0.72 \\
No. sound played & 1.06 $\pm$ 0.44 & 6.68 $\pm$ 2.34\\
Total time (sec) & 27.4 $\pm$ 11.1 & 45.8 $\pm$ 18.4 \\ 

\bottomrule
\end{tabular}
\end{table}


Table 1 further compares user behaviors between \emph{Vroom!} and \emph{TextSearch}. Specifically, for each search, average $\pm$ stand deviation values of the number of ``Go Search!'' button clicked (search trials), the number of candidate sounds played, and the total time spent are reported. 
Compared with \emph{TextSearch}, \emph{Vroom!} has significantly less search trials on average. This suggests that vocal imitation queries were generally easier to cast than keywords in our experiments. The number of sound candidates played in a search using \emph{Vroom!}, however, is much larger than that using \emph{TextSearch}. This is reasonable, as for vocal-imitation-based search the only way to make the final selection from the returned sound candidates was to listen through them, while for text-based search subjects often relied on text match. We argue, however, this advantage of text-based search would vanish as the sound library enlarges to contain multiple sounds sharing the same keywords. The overall time spent on each search in \emph{Vroom!} is significantly longer than that in \emph{TextSearch}. This can be explained by the larger number of sounds played in \emph{Vroom!} as well as the additional time spent to record and playback vocal imitations compared to typing keywords.

\section{CONCLUSIONS AND DISCUSSIONS}
\label{sec:conclusions}

This paper presented a subjective study to compare vocal-imitation-based and text-based search for sounds. We designed a search engine for each approach and an experimental framework for the study. User ratings and behavioral data collected from 20 subjects showed that vocal-imitation-based search has significant advantages over text-based search for certain categories (e.g., Synthesizers and Acoustical Instruments) of sounds in our limited sound library. Ease-of-use ratings of the vocal-imitation-based engine is also significantly higher than that of the text-based engine.


It has to be admitted that this study has some limitations. First, the sound library only contains three sound categories and each one is quite small. The former is an issue as the comparison between vocal-imitation-based search and text-based search varies much from one category to another. The latter is an issue because both search approaches would encounter difficulties when there are many similar sounds with similar keywords in a category; fine grained search methods would be needed. Second, the text-based search baseline could become stronger by enriching keyword annotations of sounds in the library. However, the lack of well annotated keywords is a common problem for user-uploaded sounds in practice. For future work, we would like to address these limitations by conducting large-scale subjective studies with a larger sound library.




\vfill\pagebreak
\bibliographystyle{IEEEtran}
\bibliography{output.bbl}
%
%
%
%
%
%
%
%
%

\end{sloppy}
\end{document}